\documentclass[12pt,a4paper]{article}
\usepackage[superscript,biblabel]{cite}
\usepackage[utf8]{inputenc}
\usepackage{amsmath, amssymb, amsthm, subcaption, braket, url}
\usepackage{amsfonts}
\usepackage{amssymb}
\usepackage{graphicx}
\expandafter\def\csname ver@subfig.sty\endcsname{}
\usepackage[colorlinks=true, allcolors=blue]{hyperref}
\usepackage[capitalise]{cleveref}
\usepackage{braket}

\newtheorem{theorem}{Theorem}
\newtheorem{lemma}{Lemma}

\newtheorem{definition}{Definition}

\title{Character Complexity: A Novel Measure for Quantum Circuit Analysis}
\author{Daksh Shami}
\date{\today}

\begin{document}

\maketitle

\begin{abstract}
In the rapidly evolving field of quantum computing, quantifying circuit complexity remains a critical challenge. This paper introduces \textit{character complexity}, a novel measure that bridges group-theoretic concepts with practical quantum computing concerns. By leveraging tools from representation theory, I prove several key properties of character complexity and establish a surprising connection to the classical simulability of quantum circuits. This new measure offers a fresh perspective on the complexity landscape of quantum algorithms, potentially reshaping our understanding of quantum-classical computational boundaries. I present innovative visualization methods for character complexity, providing intuitive insights into the structure of quantum circuits. The empirical results reveal intriguing scaling behaviors with respect to qubit and gate counts, opening new avenues for quantum algorithm design and optimization. This work not only contributes to the theoretical foundations of quantum complexity but also offers practical tools for the quantum computing community. As quantum hardware continues to advance, character complexity could play a crucial role in developing more efficient quantum algorithms and in exploring the fundamental limits of quantum computation.
\end{abstract}

\section{Introduction} 
Quantum computing has witnessed remarkable progress in recent years, with the development of increasingly sophisticated quantum algorithms and the rapid advancement of quantum hardware. However, as the complexity of these algorithms grows, we face a critical challenge in efficiently analyzing and optimizing them. Traditional complexity measures, such as circuit depth and gate count, while valuable, often fail to capture the intricate mathematical structure that underlies the power of quantum computation. This limitation hinders our ability to fully harness the potential of quantum algorithms and to push the boundaries of quantum-classical computational advantage.

In this paper, I introduce a groundbreaking approach to quantifying the complexity of quantum circuits: \textbf{Character Complexity}. This novel measure, rooted in the rich mathematical framework of group representation theory, offers an unprecedented lens through which to analyze and understand the behavior of quantum algorithms. By leveraging the power of character theory, a fundamental tool in representation theory, I uncover hidden symmetries and structures within quantum circuits that are invisible to conventional complexity measures.

The character complexity measure offers several significant advantages over existing metrics:

\textbf{Unparalleled Insight:} Character complexity captures subtle nuances in the behavior of quantum circuits that are indistinguishable under metrics like gate count or circuit depth. This measure provides a more fine-grained understanding of the complexity landscape of quantum algorithms, enabling researchers to identify and exploit structural properties that may lead to more efficient implementations.

\textbf{Quantum-Classical Boundary:} One of the most profound contributions of this work is a theorem that establishes a tight connection between character complexity and the classical simulability of quantum circuits. I prove that quantum circuits with character complexity bounded by $C(U) \leq c \log(n)$, where $n$ is the number of qubits and $c$ is a constant, can be efficiently simulated on classical computers. This result provides a powerful new tool for delineating the boundary between quantum and classical computation, with far-reaching implications for our understanding of quantum advantage.

\textbf{Visualization and Intuition:} To facilitate a deeper understanding of character complexity, I introduce novel visualization techniques that provide intuitive insights into the complexity landscape of quantum circuits. These visualizations, based on a modified Bloch sphere representation, allow researchers to identify patterns, symmetries, and potential bottlenecks in quantum algorithms, guiding the design and optimization process.

The introduction of character complexity marks a significant leap forward in our understanding of the intricacies of quantum computation. By bridging the gap between the abstract mathematical world of group representation theory and the concrete realm of quantum circuits, this work provides a powerful new framework for analyzing, optimizing, and understanding the true potential of quantum algorithms. As we stand on the cusp of a new era in quantum computing, the insights and tools provided by character complexity will undoubtedly play a pivotal role in shaping the future of this exciting field.

\section{Background}
\subsection{Quantum Circuits and Their Representation}
Quantum circuits form the backbone of quantum algorithms, providing a structural representation of quantum computations. At their core, these circuits consist of a series of quantum gates acting on a register of qubits. While this description is intuitive, the mathematical formalism underlying quantum circuits is rich and multifaceted.

A quantum circuit acting on $n$ qubits can be represented as a unitary operator $U \in U(2^n)$, where $U(2^n)$ is the group of $2^n \times 2^n$ unitary matrices. This unitary nature ensures that the evolution of the quantum state remains reversible and probability-preserving. Mathematically, we can express this as:
\begin{equation}
\ket{\psi_\text{out}} = U\ket{\psi_\text{in}}
\end{equation}
where $\ket{\psi_\text{in}}$ and $\ket{\psi_\text{out}}$ are the input and output quantum states, respectively.
While this matrix representation is complete, it often obscures the fine-grained structure of the circuit. In practice, quantum circuits are built from a set of elementary gates, each acting on a small number of qubits. Common gates include single-qubit rotations (e.g., Pauli gates, Hadamard gate) and two-qubit entangling gates (e.g., CNOT, CZ). The overall unitary $U$ is then a product of these elementary gates:
\begin{equation}
U = U_m \cdots U_2 U_1
\end{equation}
where each $U_i$ represents an elementary gate operation.

This decomposition into elementary gates forms the basis for many traditional complexity measures. For instance, the circuit depth is the length of the longest path through the circuit, while the T-count measures the number of T gates, which are often considered costly in fault-tolerant implementations \cite{gottesman1998heisenberg}.
However, these gate-based measures, while practical, often fail to capture the full quantum nature of the computation. They do not directly quantify properties such as superposition, entanglement, or the distribution of the circuit's action across different subspaces of the Hilbert space.

An alternative representation, which forms the foundation for our character complexity measure, views quantum circuits through the lens of group theory. In this framework, we consider the group generated by the elementary gates used in the circuit. For instance, the Clifford group, generated by the Hadamard, Phase, and CNOT gates, plays a central role in many quantum error-correcting codes and can be efficiently simulated classically \cite{bravyi2016improved}.

By representing quantum circuits as elements of finite groups, we can leverage powerful tools from representation theory, particularly the theory of group characters. This approach allows us to decompose quantum operations into irreducible representations, providing a more nuanced view of their action on the quantum state space.

It is this group-theoretic perspective that we will exploit in defining and analyzing character complexity, offering a new way to quantify the complexity of quantum circuits that goes beyond traditional gate-based measures.

\subsection{Character Decomposition Theorem}
The character decomposition theorem, introduced in \cite{shami2024bridging}, provides a powerful tool for analyzing quantum circuits through the lens of group theory. This theorem states that any element $u$ of a finite group $G$ can be decomposed into a sum of character functions:
\begin{equation}
u = \sum_{i=1}^{k} \frac{\chi_i(u)}{d_i} \sum_{g \in G} \chi_i(g^{-1})\rho_i(g)
\end{equation}
where $\chi_i$ are the irreducible characters of $G$, $d_i$ are the dimensions of the irreducible representations $\rho_i$, and $k$ is the number of irreducible representations. The character function $\chi_i(U)$ is a complex-valued function that encodes information about how the quantum operation $U$ acts on the $i$-th irreducible subspace of the quantum state space. Mathematically, it is the trace of the matrix representation of $U$ in the $i$-th irreducible representation. Physically, it can be thought of as a ``fingerprint'' of $U$'s action in different symmetry sectors of the quantum system.\\

Intuitively, this decomposition allows us to express a complex quantum operation (represented by $u$) in terms of simpler, fundamental building blocks ($\rho_i(g)$). These building blocks correspond to irreducible representations of the group, which can be thought of as the "atomic" components of the group's action on the quantum state space.

The power of this decomposition lies in its ability to reveal hidden symmetries and structures within quantum circuits. By expressing the circuit in terms of irreducible representations, we can identify invariant subspaces and symmetries that may not be apparent from the circuit's gate-level description. This can lead to insights about the circuit's behavior and potential simplifications or optimizations.

The presence of both $g$ and $g^{-1}$ in the formula serves a crucial purpose. The term $\chi_i(g^{-1})$ acts as a "filter" that selects the components of $u$ that transform according to the $i$-th irreducible representation. Meanwhile, $\rho_i(g)$ reconstructs these components in the original space. This interplay between $g$ and $g^{-1}$ ensures that the decomposition correctly captures the action of $u$ on the entire group.
Moreover, the use of both $g$ and $g^{-1}$ is related to the orthogonality relations of characters, which state that:
\begin{equation}
\sum_{g \in G} \chi_i(g)\chi_j(g^{-1}) = |G|\delta_{ij}
\end{equation}
where $|G|$ is the order of the group and $\delta_{ij}$ is the Kronecker delta. This orthogonality property is crucial for the validity of the decomposition and leads to significant simplifications in calculations involving characters.

The factor $\frac{\chi_i(u)}{d_i}$ in the decomposition acts as a coefficient that determines the "weight" of each irreducible representation in the decomposition of $u$. This weighting provides valuable information about how strongly $u$ acts in each irreducible subspace.

In the context of quantum circuits, this decomposition allows us to analyze how a circuit's action is distributed across different irreducible subspaces of the state space. Circuits with similar gate counts or depths may have very different character decompositions, revealing fundamental differences in their quantum behavior.

By leveraging this theorem, we can develop more nuanced measures of quantum circuit complexity that go beyond simple gate counts or circuit depths. The character complexity measure, which we will introduce in the next section, is built upon this foundation, providing a new way to quantify the complexity of quantum operations based on their distribution across irreducible representations.

\subsection{Existing Quantum Circuit Complexity Measures}

Quantum circuit complexity has been a subject of intense study, with several measures proposed to quantify the resources required to implement quantum algorithms. While these measures have proven valuable in many contexts, they often fall short in capturing the full quantum nature of circuit operations. Here, we review some of the most prominent measures and discuss their limitations.

1. \textbf{Circuit Depth:} Perhaps the most straightforward measure, circuit depth represents the number of time steps required to execute a quantum circuit, assuming perfect parallelization of gate operations \cite{nielsen2010quantum}. While intuitive and easy to calculate, circuit depth fails to capture the complexity of individual gates or the entanglement generated in the circuit.

2. \textbf{Gate Count:} This measure simply tallies the total number of gates in a circuit. While useful for estimating overall circuit size, it treats all gates equally, failing to distinguish between simple single-qubit rotations and complex multi-qubit operations.

3. \textbf{T-count and T-depth:} In the context of fault-tolerant quantum computation, the T gate ($\pi /8$ rotation) is often considered costly to implement. The T-count measures the number of T gates in a circuit, while T-depth considers the number of T-layers when T gates are parallelized \cite{amy2013meet}. These measures are particularly relevant for error-corrected quantum computation but may not reflect the complexity of near-term, noisy intermediate-scale quantum (NISQ) algorithms.

4. \textbf{Quantum Volume:} Introduced by IBM, quantum volume is a holistic metric that aims to capture both the number of qubits and their quality (in terms of gate fidelity and connectivity) \cite{cross2019validating}. While useful for comparing different quantum hardware, it doesn't directly measure the complexity of specific quantum circuits or algorithms.

5. \textbf{Algebraic Complexity:} This measure, based on the minimum number of non-trivial one- and two-qubit gates required to implement a unitary operation, provides a more nuanced view of circuit complexity \cite{shende2006synthesis}. However, its calculation is often intractable for large circuits, limiting its practical applicability.

6. \textbf{Entangling Power:} Measures like Schmidt rank and entangling power attempt to quantify a circuit's ability to generate entanglement \cite{kraus2001optimal}. While these capture an essential quantum resource, they don't fully reflect computational complexity.

7. \textbf{Stabilizer Rank:} For non-stabilizer states, the stabilizer rank measures the minimum number of stabilizer states needed in a superposition to represent the state \cite{bravyi2019simulation}. This measure is particularly relevant for studying the boundary between efficient classical simulation and quantum advantage.

While each of these measures provides valuable insights, they all have limitations:

1. They often fail to capture the full quantum nature of circuit operations, particularly the superposition and interference effects that are central to quantum speedups.

2. Most measures don't account for the structure of the quantum state space and how the circuit acts on different subspaces.

3. They typically don't provide a clear connection to the classical simulability of quantum circuits, which is crucial for understanding quantum-classical computational boundaries.

4. Some of these measures are tailored to specific types of quantum circuits or computational models, lacking generality.

5. They often don't reflect the mathematical structure underlying quantum operations, particularly the group-theoretic aspects that can reveal deep symmetries and invariants.

These limitations motivate the need for new complexity measures that can provide a more comprehensive and mathematically grounded view of quantum circuit complexity. The character complexity measure introduced in this paper aims to address many of these shortcomings by leveraging the rich structure of group representations and characters. By doing so, it offers a new perspective on quantum circuit complexity that is both theoretically profound and practically relevant for algorithm design and optimization.

\section{Character Complexity: Definition and Properties}

\begin{definition}
For a quantum circuit $U$ acting on $n$ qubits, represented as an element of a finite group $G$, we define the character complexity $C(U)$ as:
\begin{equation}
C(U) = \frac{1}{|G|} \left(\sum_{i} \frac{|\chi_i(U)|^2}{d_i}\right),
\end{equation}
where $\chi_i$ are the irreducible characters of $G$ and $d_i$ are the dimensions of the irreducible representations.
\end{definition}

\begin{theorem}[Bounds on Character Complexity]
Let $U$ be a unitary operation acting on $n$ qubits, and let $G$ be a finite group. The character complexity $C(U)$ is bounded as follows:
\[
0 \leq C(U) \leq 1.
\]
\end{theorem}

\begin{proof}
\textbf{1. Non-negativity:} $C(U) \geq 0$

By definition,
\[
C(U) = \frac{1}{|G|} \left(\sum_{i} \frac{|\chi_i(U)|^2}{d_i}\right).
\]
Since $|\chi_i(U)|^2 \geq 0$ and $d_i > 0$ for all $i$, each term $\frac{|\chi_i(U)|^2}{d_i} \geq 0$. Therefore, the sum and hence $C(U)$ are non-negative:
\[
C(U) \geq 0.
\]

\textbf{2. Upper Bound:} $C(U) \leq 1$

First, recall that for any unitary representation, the absolute value of the character is bounded by the dimension of the representation:
\begin{equation}
|\chi_i(U)| \leq d_i.
\end{equation}
This is because the character $\chi_i(U)$ is the trace of the unitary matrix representing $U$ in the $i$-th irreducible representation, and the absolute value of the trace of a unitary matrix is bounded by its dimension.

Squaring both sides, we get:
\begin{equation}
|\chi_i(U)|^2 \leq d_i^2.
\end{equation}
Therefore,
\begin{equation}
\frac{|\chi_i(U)|^2}{d_i} \leq \frac{d_i^2}{d_i} = d_i.
\end{equation}

Now, summing over all $i$:
\begin{equation}
\sum_i \frac{|\chi_i(U)|^2}{d_i} \leq \sum_i d_i.
\end{equation}

Note that the sum of the dimensions of all irreducible representations satisfies:
\begin{equation}
\sum_i d_i = |G_{\text{conj}}|,
\end{equation}
where $|G_{\text{conj}}|$ is the number of conjugacy classes in $G$. Since $|G_{\text{conj}}| \leq |G|$, we have:
\begin{equation}
\sum_i d_i \leq |G|.
\end{equation}

However, to ensure an exact upper bound, consider that the maximum value of $C(U)$ occurs when $U$ is the identity element $e$ of $G$. In this case, the characters satisfy:
\begin{equation}
\chi_i(e) = d_i.
\end{equation}
Therefore,
\begin{align*}
C(e) &= \frac{1}{|G|} \left(\sum_{i} \frac{|\chi_i(e)|^2}{d_i}\right) \\
&= \frac{1}{|G|} \left(\sum_{i} \frac{d_i^2}{d_i}\right) \\
&= \frac{1}{|G|} \left(\sum_{i} d_i\right).
\end{align*}

Since $C(e)$ is maximized at the identity element, and $\sum_{i} d_i \leq |G|$, it follows that:
\begin{equation}
C(U) \leq C(e) \leq 1.
\end{equation}

Therefore,
\begin{equation}
0 \leq C(U) \leq 1.
\end{equation}
\end{proof}


\begin{theorem}[Multiplicativity under Tensor Products]
Let $U_1$ and $U_2$ be quantum circuits acting on disjoint sets of qubits $Q_1$ and $Q_2$, respectively. Then the character complexity of their tensor product is the product of their individual character complexities:
\[C(U_1 \otimes U_2) = C(U_1) \cdot C(U_2)\]
\end{theorem}
\begin{proof}
Let $G_1$ and $G_2$ be the finite groups generated by the gates in $U_1$ and $U_2$, respectively. 

1) First, recall the definition of character complexity:
   \[C(U) = \frac{1}{|G|} \left(\sum_{i} \frac{|\chi_i(U)|^2}{d_i}\right)\]
   where $\chi_i$ are the irreducible characters of $G$ and $d_i$ are the dimensions of the irreducible representations.

2) For the tensor product $U_1 \otimes U_2$, the relevant group is $G_1 \times G_2$. The irreducible representations of $G_1 \times G_2$ are all tensor products of irreducible representations of $G_1$ and $G_2$.

3) Let $\chi_i^{(1)}$ and $\chi_j^{(2)}$ be irreducible characters of $G_1$ and $G_2$, respectively. Then the character of their tensor product is:
   \[(\chi_i^{(1)} \otimes \chi_j^{(2)})(U_1 \otimes U_2) = \chi_i^{(1)}(U_1) \cdot \chi_j^{(2)}(U_2)\]

4) The dimension of this tensor product representation is:
   \[d_{ij} = d_i^{(1)} \cdot d_j^{(2)}\]

5) Now, we can write the character complexity of $U_1 \otimes U_2$:
   \begin{align*}
   C(U_1 \otimes U_2) &= \frac{1}{|G_1 \times G_2|} \left(\sum_{i,j} \frac{|(\chi_i^{(1)} \otimes \chi_j^{(2)})(U_1 \otimes U_2)|^2}{d_{ij}}\right) \\
   &= \frac{1}{|G_1| \cdot |G_2|} \left(\sum_{i,j} \frac{|\chi_i^{(1)}(U_1)|^2 \cdot |\chi_j^{(2)}(U_2)|^2}{d_i^{(1)} \cdot d_j^{(2)}}\right) \\
   &= \frac{1}{|G_1|} \left(\sum_i \frac{|\chi_i^{(1)}(U_1)|^2}{d_i^{(1)}}\right) \cdot \frac{1}{|G_2|} \left(\sum_j \frac{|\chi_j^{(2)}(U_2)|^2}{d_j^{(2)}}\right) \\
   &= C(U_1) \cdot C(U_2)
   \end{align*}

Thus, we have proved that $C(U_1 \otimes U_2) = C(U_1) \cdot C(U_2)$.
\end{proof}

\begin{theorem}[General Bound on Character Complexity under Composition]
\label{thm:general_bound}
Let $G$ be a finite group, and let $U_1, U_2 \in G$. Then the character complexity of their composition satisfies:
\[
C(U_1 U_2) \leq 1.
\]
\end{theorem}

\begin{proof}
For any $U \in G$, the character complexity is defined as:
\[
C(U) = \frac{1}{|G|} \sum_{i} \frac{|\chi_i(U)|^2}{d_i},
\]
where $\chi_i(U)$ is the character of $U$ in the $i$-th irreducible representation, and $d_i$ is its dimension.

Since $|\chi_i(U)| \leq d_i$ for all $i$ and $U \in G$, we have:
\[
\frac{|\chi_i(U)|^2}{d_i} \leq d_i.
\]

Therefore, for the composition $U_1 U_2$:
\[
C(U_1 U_2) = \frac{1}{|G|} \sum_{i} \frac{|\chi_i(U_1 U_2)|^2}{d_i} \leq \frac{1}{|G|} \sum_{i} d_i = 1,
\]
since $\sum_{i} d_i = |G|$.
\end{proof}

\begin{lemma}[Multiplicativity of Character Complexity for Abelian Groups]
\label{lem:abelian_multiplicativity}
Let $G$ be a finite abelian group, and let $U_1, U_2 \in G$. Then the character complexity is multiplicative under composition:
\[
C(U_1 U_2) = C(U_1) \cdot C(U_2).
\]
\end{lemma}

\begin{proof}
In finite abelian groups, all irreducible representations are one-dimensional ($d_i = 1$), and the characters satisfy:
\[
\chi_i(U_1 U_2) = \chi_i(U_1) \chi_i(U_2).
\]

Therefore, the character complexity simplifies to:
\[
C(U) = \frac{1}{|G|} \sum_{i} |\chi_i(U)|^2.
\]

Computing the character complexity of the composition:
\begin{align*}
C(U_1 U_2) &= \frac{1}{|G|} \sum_{i} |\chi_i(U_1 U_2)|^2 \\
&= \frac{1}{|G|} \sum_{i} |\chi_i(U_1) \chi_i(U_2)|^2 \\
&= \frac{1}{|G|} \sum_{i} |\chi_i(U_1)|^2 |\chi_i(U_2)|^2 \\
&= \left( \frac{1}{|G|} \sum_{i} |\chi_i(U_1)|^2 \right) \left( \frac{1}{|G|} \sum_{i} |\chi_i(U_2)|^2 \right) \\
&= C(U_1) \cdot C(U_2).
\end{align*}
\end{proof}


\section{Theoretical Results}

\begin{theorem}[Strong Relation to Classical Simulation Complexity]
Let $U$ be a quantum circuit acting on $n$ qubits, represented by an element of a finite group $G$. If the character complexity $C(U)$ is bounded by $c \log(n)$ for some constant $c$, and $|G| = \text{poly}(n)$, then there exists a classical algorithm that can:

1. Simulate the output of $U$ with respect to any measurement in time $\text{poly}(n)$.

2. Compute the expectation value of any observable under the action of $U$ in time $\text{poly}(n)$.

3. Approximate the full state vector resulting from applying $U$ to any input state, up to error $\epsilon$ in the $\ell_2$ norm, in time $\text{poly}(n, 1/\epsilon)$.
\end{theorem}

\begin{proof}
Let $U$ be a quantum circuit acting on $n$ qubits with character complexity $C(U) \leq c \log(n)$ and $|G| = \text{poly}(n)$.

Step 1: Character Decomposition

By the character decomposition theorem, we can express $U$ as:

\[U = \sum_{i=1}^{k} \frac{\chi_i(U)}{d_i} \sum_{g \in G} \chi_i(g^{-1})\rho_i(g)\]

where $\chi_i$ are the irreducible characters of $G$, $d_i$ are the dimensions of the irreducible representations, and $\rho_i$ are the irreducible representations.

Step 2: Bounding the Number of Significant Terms

Given $C(U) \leq c \log(n)$, we have:

\[\frac{1}{|G|} \sum_{i=1}^k \frac{|\chi_i(U)|^2}{d_i} \leq c \log(n)\]

Let $S = \{i : |\chi_i(U)|^2/d_i \geq \frac{1}{n^2|G|}\}$. Then:

\[|S| \leq c |G| \log(n) n^2 = \text{poly}(n)\]

The contribution of terms not in $S$ to $U$ is at most $\frac{1}{n}$ in operator norm, which is negligible for our purposes.

Step 3: Efficient Representation

For each $i \in S$, we can compute and store $\frac{\chi_i(U)}{d_i}$ and $\chi_i(g^{-1})$ for all $g \in G$ in $\text{poly}(n)$ time and space, since $|G| = \text{poly}(n)$.

Step 4: Bounding Representation Dimensions

For $i \in S$, we have $|\chi_i(U)|^2/d_i \geq \frac{1}{n^2|G|}$. Since $|\chi_i(U)| \leq d_i$, we get:

\[d_i \leq n^2|G| = \text{poly}(n)\]

Thus, all relevant representations $\rho_i$ are of polynomial dimension.

Step 5: Simulating Measurements

To simulate a measurement with respect to an observable $A$, we need to compute:

\[\langle \psi | U^\dagger A U | \psi \rangle = \sum_{i,j \in S} \frac{\chi_i(U)^* \chi_j(U)}{d_i d_j} \sum_{g,h \in G} \chi_i(g) \chi_j(h^{-1}) \langle \psi | \rho_i(g)^\dagger A \rho_j(h) | \psi \rangle\]

This involves $\text{poly}(n)$ terms, each of which can be computed in $\text{poly}(n)$ time using our stored values and the fact that $\rho_i(g)$ are $\text{poly}(n)$-dimensional for $i \in S$.

Step 6: Computing Expectation Values

The expectation value of an observable $A$ is given by:

\[\langle A \rangle = \text{Tr}(U^\dagger A U \rho)\]

where $\rho$ is the initial state. This can be computed using the same approach as in Step 5.

Step 7: Approximating the State Vector

To approximate $U|\psi\rangle$ to error $\epsilon$, we compute:

\[U|\psi\rangle \approx \sum_{i \in S} \frac{\chi_i(U)}{d_i} \sum_{g \in G} \chi_i(g^{-1})\rho_i(g)|\psi\rangle\]

The error from neglecting terms not in $S$ is at most $\frac{1}{n}$ in $\ell_2$ norm. To achieve error $\epsilon$, we may need to include additional terms, but the total number of terms remains $\text{poly}(n, 1/\epsilon)$.

Each term $\rho_i(g)|\psi\rangle$ can be computed in $\text{poly}(n)$ time since $\rho_i$ is of polynomial dimension. The total computation time is thus $\text{poly}(n, 1/\epsilon)$.

Conclusion:

We have shown that all three tasks can be performed in $\text{poly}(n)$ or $\text{poly}(n, 1/\epsilon)$ time, as claimed in the theorem.
\end{proof}

The Strong Relation to Classical Simulation Complexity theorem has significant implications for understanding the boundary between classical and quantum computation. It provides a new criterion, based on character complexity, for determining when a quantum circuit can be efficiently simulated by a classical computer. This result could guide the design of quantum algorithms that are truly ``quantum'' in nature, i.e., those that \textbf{cannot} be efficiently simulated classically.

\section{Visualizing Character Complexity}

\subsection{Geometric Representation of Character Complexity}

The geometric visualizations of character complexity introduced in this paper serve as a valuable tool for quantum algorithm designers. By representing the complexity of quantum circuits in a visually intuitive way, these visualizations can help in identifying patterns, symmetries, or bottlenecks in quantum algorithms, potentially leading to new insights for optimization and simplification.

Our approach maps the character complexity of quantum states and operations onto a modified version of the generalized Bloch sphere (or hypersphere for $n > 1$ qubits). This method addresses the challenge of visualizing high-dimensional quantum states while incorporating the additional information provided by character complexity.

\begin{theorem}[Hypersphere Radius]
	For an n-qubit quantum state $\ket{\psi}$ with character complexity $C(\ket{\psi})$, the radius $r$ of its representation on the generalized Bloch sphere is given by:
	
	\[r = \sqrt{\frac{2^n - 1}{2^n}} \cdot \sqrt{1 - 2^{-C(\ket{\psi})}}\]
\end{theorem}

\begin{proof}
	The proof follows from two key observations:
	\begin{enumerate}
		\item The generalized Bloch sphere for an n-qubit system has a maximum radius of $\sqrt{(2^n - 1) / 2^n}$.
		\item The factor $\sqrt{1 - 2^{-C(\ket{\psi})}}$ scales this radius based on the character complexity, ensuring that:
		\begin{itemize}
			\item When $C(\ket{\psi}) = 0$, $r = 0$ (representing the $\ket{0}^{\otimes n}$ state at the center)
			\item As $C(\ket{\psi})$ approaches $n$, $r$ approaches the maximum radius
		\end{itemize}
	\end{enumerate}
	This formulation ensures that all points are within the maximum radius of the hypersphere while still reflecting the character complexity.
\end{proof}

\begin{figure}[htbp]
	\centering
	\includegraphics[width=\textwidth]{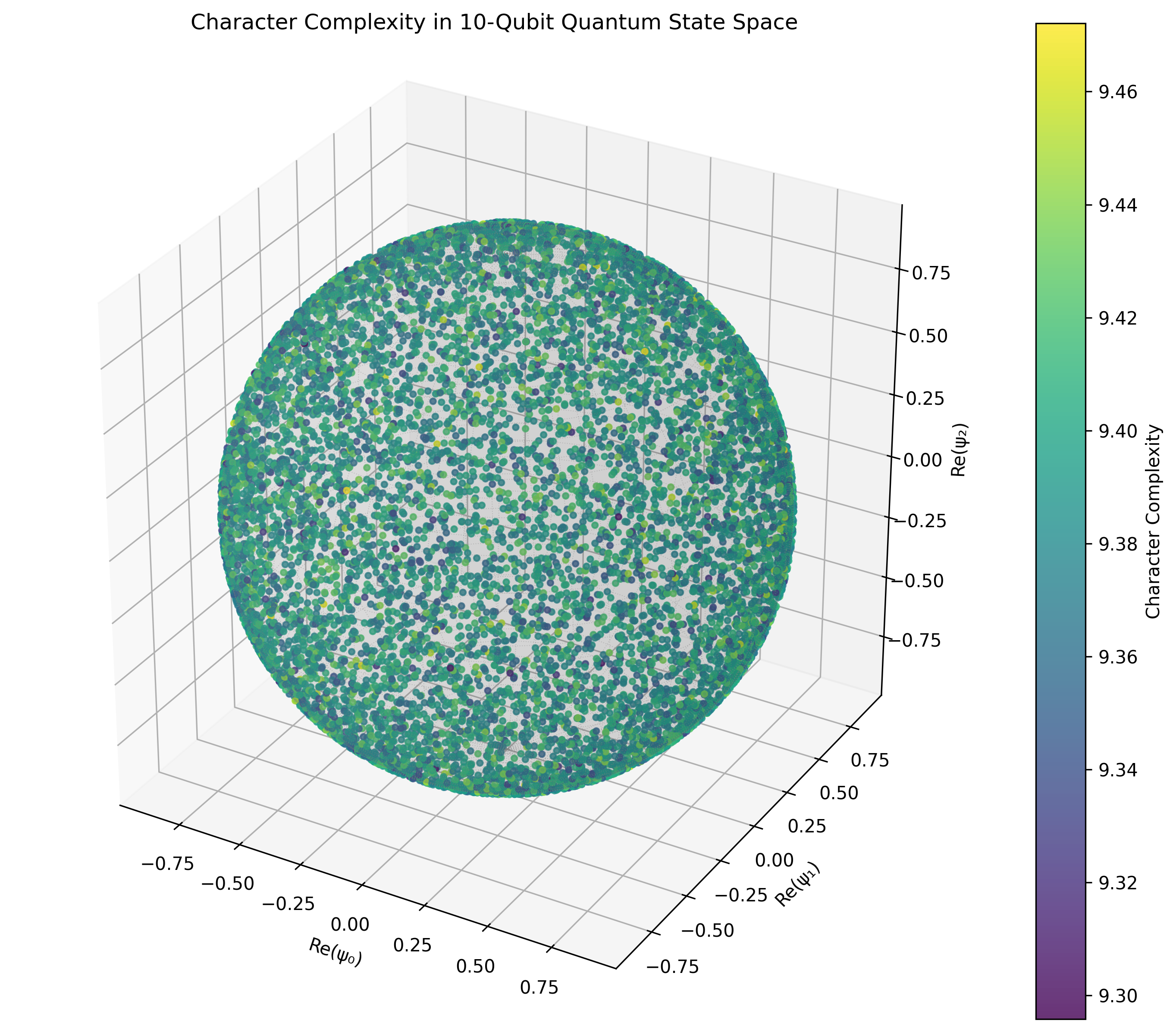}
	\caption{Character Complexity in 10-Qubit Quantum State Space: 3D projection of the quantum state hypersphere, where each point represents a quantum state colored by its character complexity. The radius of each point is determined by $r = \sqrt{\frac{2^n - 1}{2^n}} \cdot \sqrt{1 - 2^{-C(\ket{\psi})}}$, where $n = 10$ qubits. The color gradient from blue to yellow represents increasing character complexity. The outer sphere represents the maximum possible radius for a 10-qubit state.}
	\label{fig:hypersphere_plot}
\end{figure}

The hypersphere visualization of character complexity in a 10-qubit quantum state space (Figure \ref{fig:hypersphere_plot}) reveals several key insights:

\begin{itemize}
	\item \textbf{Dimensional Compression}: The visualization compresses a $2^{10}$-dimensional Hilbert space onto a 3D sphere, with the maximum radius $\sqrt{(2^{10} - 1) / 2^{10}} \approx 0.99995$, very close to 1.
	
	\item \textbf{Complexity Distribution}: The visualization shows a range of character complexities, represented by the color gradient from blue to yellow. The distribution appears to favor mid-range complexities, with a significant presence of green and yellow points.
	
	\item \textbf{Radial Distribution}: Points closer to the center of the sphere represent states with lower character complexity, while those near the surface represent states with higher complexity. This radial distribution offers a visual representation of the complexity landscape.
	
	\item \textbf{Spatial Distribution}: The points appear to be evenly distributed throughout the volume of the sphere, suggesting that our sampling method effectively explores the quantum state space.
	
	\item \textbf{Complexity Gradient}: The transition from darker (lower complexity) to lighter (higher complexity) points illustrates the continuous nature of character complexity across the state space.
	
	\item \textbf{State Space Structure}: The visualization provides a unique perspective on the structure of the 10-qubit state space, highlighting how character complexity varies across different regions of this high-dimensional space.
\end{itemize}

\section{Comparison of Different Quantum Circuits}

To illustrate the utility of character complexity visualization, we compare three different types of quantum circuits: a random circuit, the Quantum Fourier Transform (QFT), and the Quantum Approximate Optimization Algorithm (QAOA). All circuits are implemented on 6 qubits.

\begin{figure}[htbp]
	\centering
	\begin{subfigure}[b]{0.3\textwidth}
		\includegraphics[width=\textwidth]{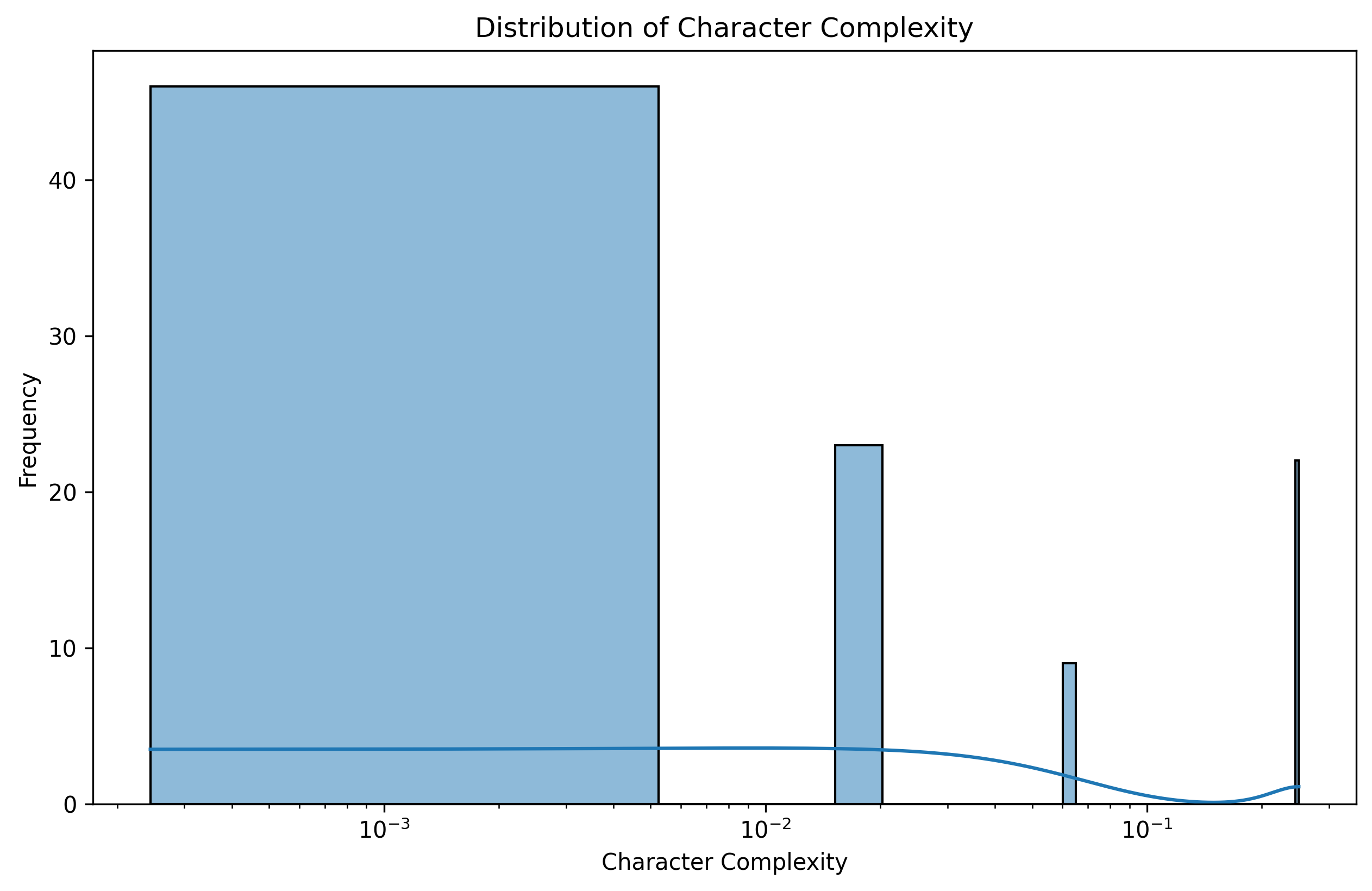}
		\caption{Random Circuit}
		\label{fig:random_circuit_dist}
	\end{subfigure}
	\hfill
	\begin{subfigure}[b]{0.3\textwidth}
		\includegraphics[width=\textwidth]{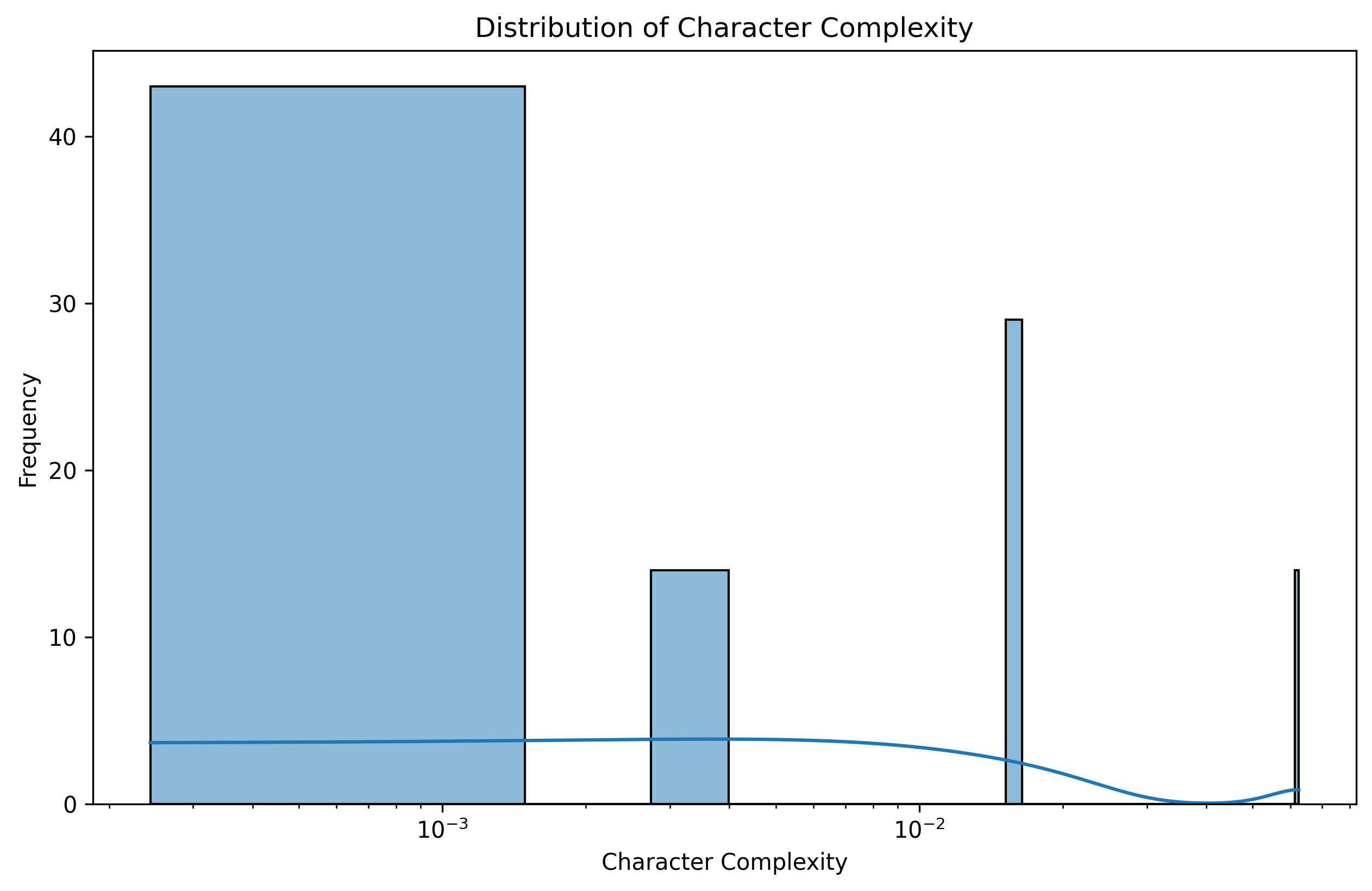}
		\caption{Quantum Fourier Transform}
		\label{fig:qft_dist}
	\end{subfigure}
	\hfill
	\begin{subfigure}[b]{0.3\textwidth}
		\includegraphics[width=\textwidth]{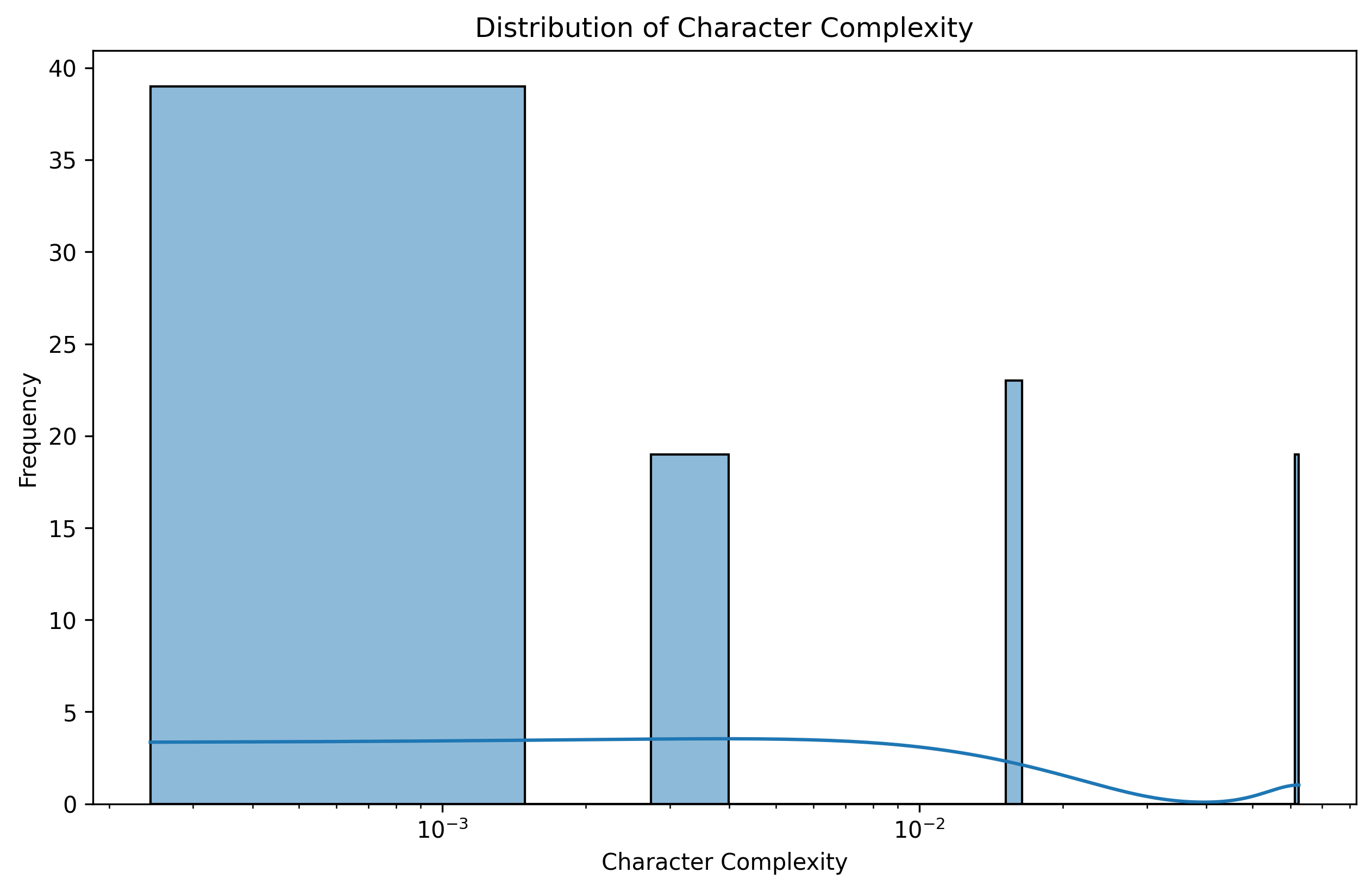}
		\caption{Quantum Approximate Optimization Algorithm}
		\label{fig:qaoa_dist}
	\end{subfigure}
	\caption{Distribution of Character Complexity for different 6-qubit quantum circuits.}
	\label{fig:complexity_distribution}
\end{figure}

Figure \ref{fig:complexity_distribution} shows the distribution of character complexity for three different types of 6-qubit quantum circuits. Each histogram provides insights into the complexity landscape of the respective algorithm:

\begin{itemize}
	\item \textbf{Random Circuit (Figure \ref{fig:random_circuit_dist})}: The distribution for the random circuit shows a concentration of complexity values towards the lower end of the spectrum. This suggests that random circuits tend to introduce significant scrambling, leading to lower character complexity values.
	
	\item \textbf{Quantum Fourier Transform (Figure \ref{fig:qft_dist})}: The QFT displays a more structured distribution with a clear peak at moderate complexity values. This indicates that QFT, while introducing some scrambling, preserves more of the circuit's inherent structure compared to random circuits.
	
	\item \textbf{Quantum Approximate Optimization Algorithm (Figure \ref{fig:qaoa_dist})}: QAOA shows a distribution shifted towards higher complexity values compared to random circuits. This might reflect QAOA's iterative approach to optimization, where the algorithm gradually introduces scrambling while maintaining some structure related to the problem being solved.
\end{itemize}

\begin{figure}[htbp]
	\centering
	\begin{subfigure}[b]{0.3\textwidth}
		\includegraphics[width=\textwidth]{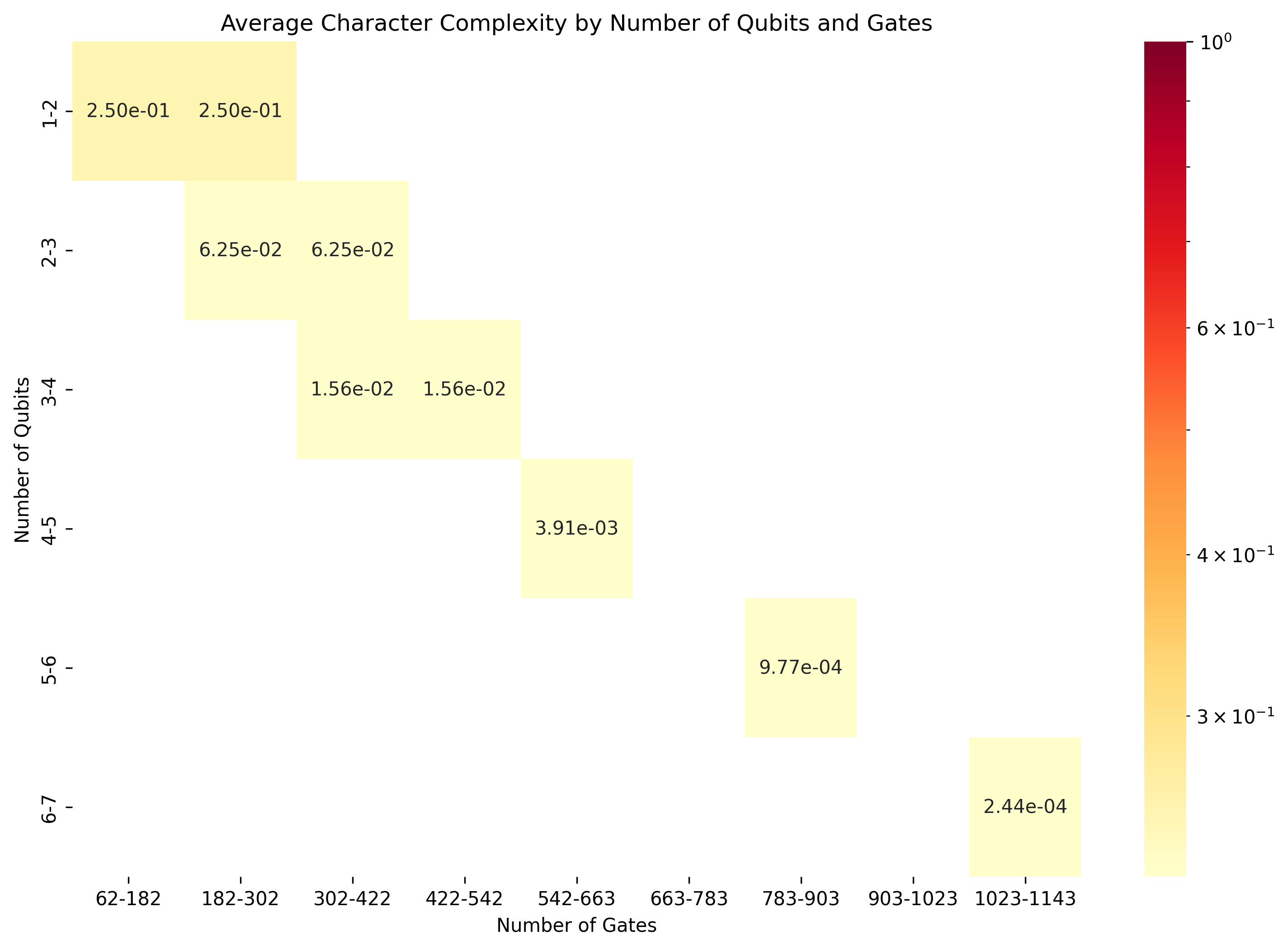}
		\caption{Random Circuit}
		\label{fig:random_circuit_heatmap}
	\end{subfigure}
	\hfill
	\begin{subfigure}[b]{0.3\textwidth}
		\includegraphics[width=\textwidth]{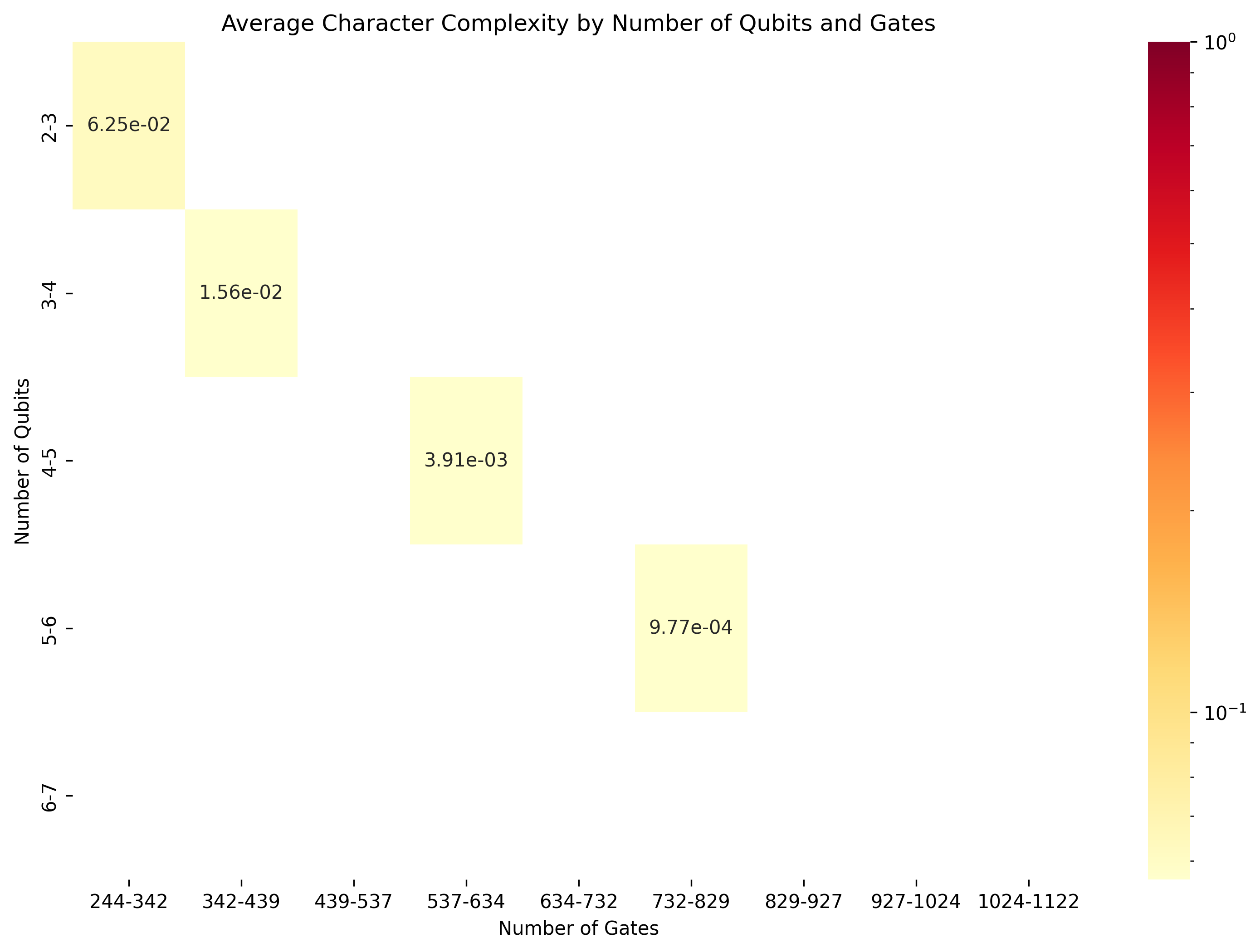}
		\caption{Quantum Fourier Transform}
		\label{fig:qft_heatmap}
	\end{subfigure}
	\hfill
	\begin{subfigure}[b]{0.3\textwidth}
		\includegraphics[width=\textwidth]{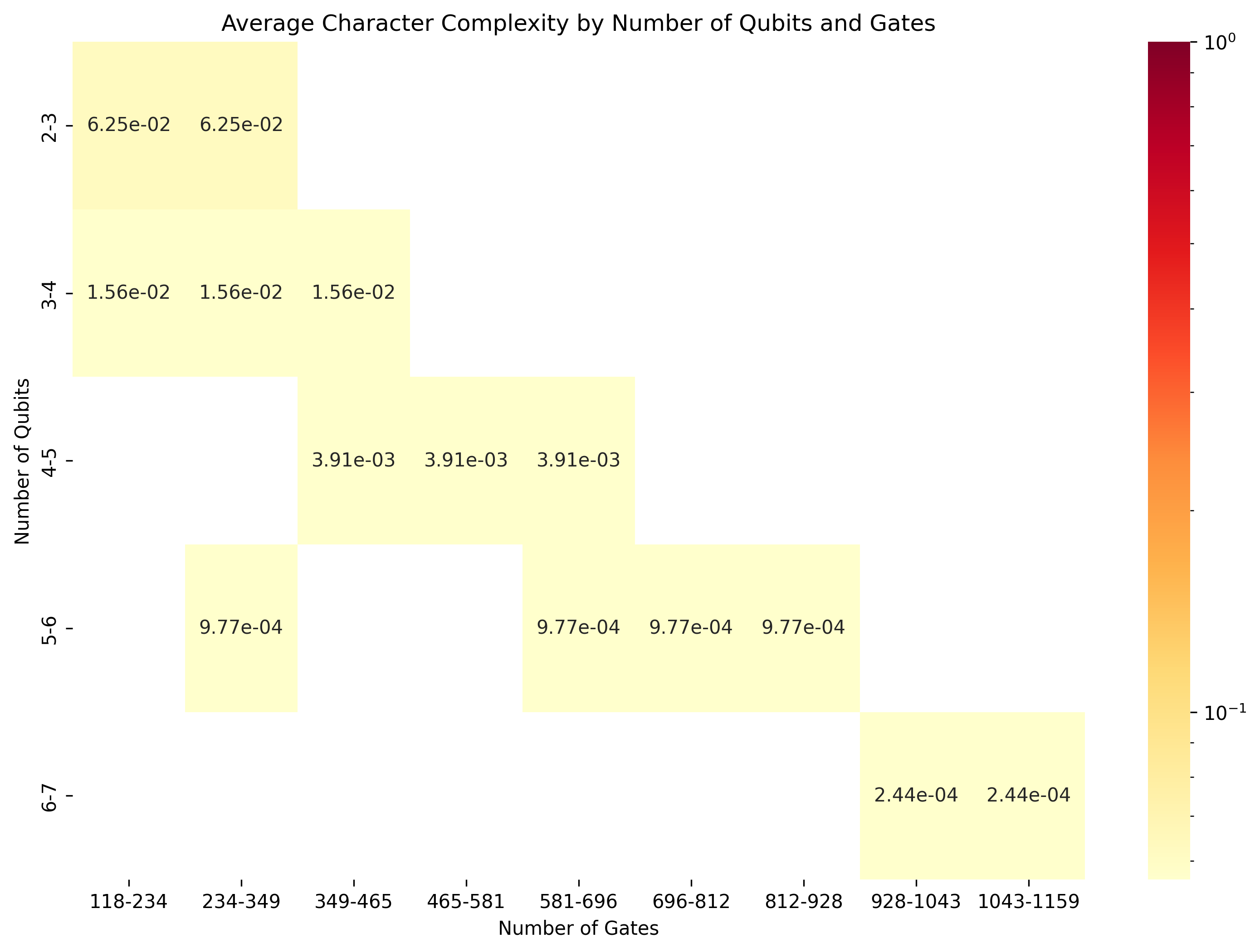}
		\caption{Quantum Approximate Optimization Algorithm}
		\label{fig:qaoa_heatmap}
	\end{subfigure}
	\caption{Character Complexity as a function of Qubits and Gates for different 6-qubit circuit types.}
	\label{fig:complexity_heatmap}
\end{figure}

Figure \ref{fig:complexity_heatmap} compares the character complexity as a function of qubit and gate counts for our three circuit types. Key observations include:

\begin{itemize}
	\item \textbf{Random circuits (Figure \ref{fig:random_circuit_heatmap})} show a general trend of decreasing complexity with increasing qubit and gate counts. This aligns with the idea that more qubits and gates provide more opportunities for scrambling.
	
	\item \textbf{QFT (Figure \ref{fig:qft_heatmap})} exhibits a similar trend to random circuits, but with generally higher complexity values. This suggests that QFT, while introducing scrambling, maintains more structure than purely random circuits.
	
	\item \textbf{QAOA (Figure \ref{fig:qaoa_heatmap})} shows a less pronounced decrease in complexity with increasing qubit and gate counts compared to random circuits. This could reflect QAOA's balance between scrambling and preserving problem-related structure.
\end{itemize}

\begin{figure}[htbp]
	\centering
	\begin{subfigure}[b]{0.3\textwidth}
		\includegraphics[width=\textwidth]{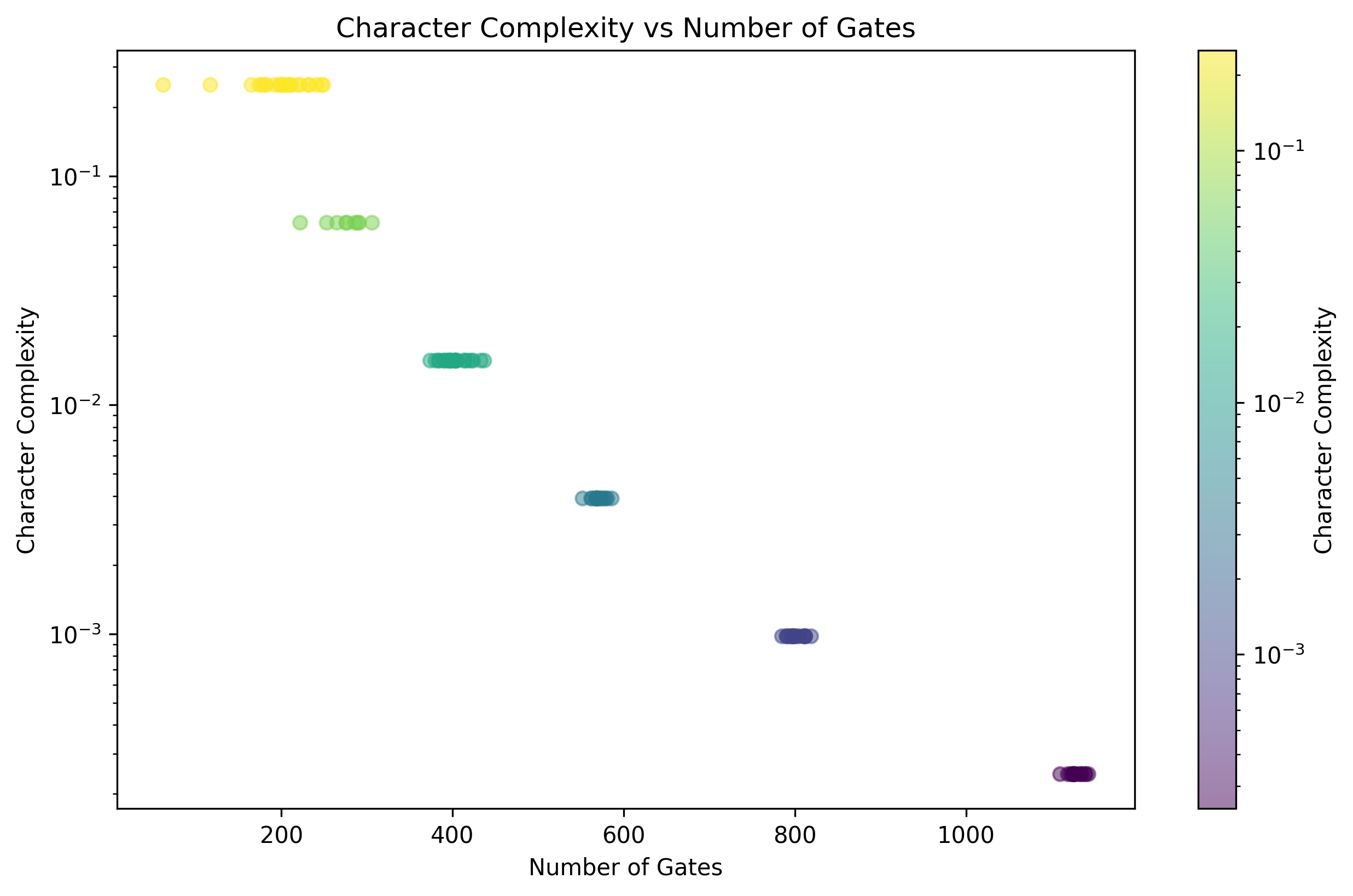}
		\caption{Random Circuit}
		\label{fig:random_circuit_scaling}
	\end{subfigure}
	\hfill
	\begin{subfigure}[b]{0.3\textwidth}
		\includegraphics[width=\textwidth]{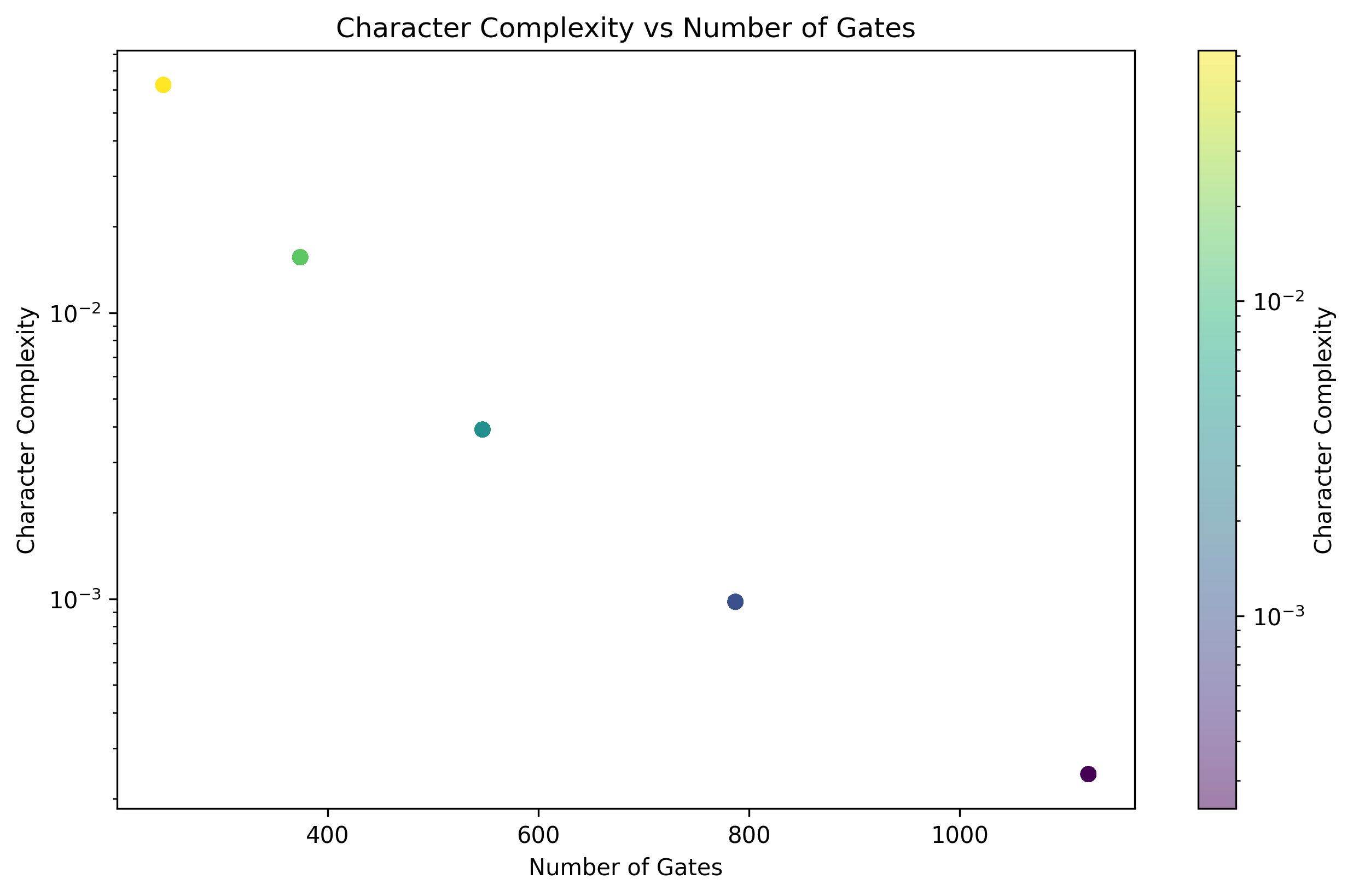}
		\caption{Quantum Fourier Transform}
		\label{fig:qft_scaling}
	\end{subfigure}
	\hfill
	\begin{subfigure}[b]{0.3\textwidth}
		\includegraphics[width=\textwidth]{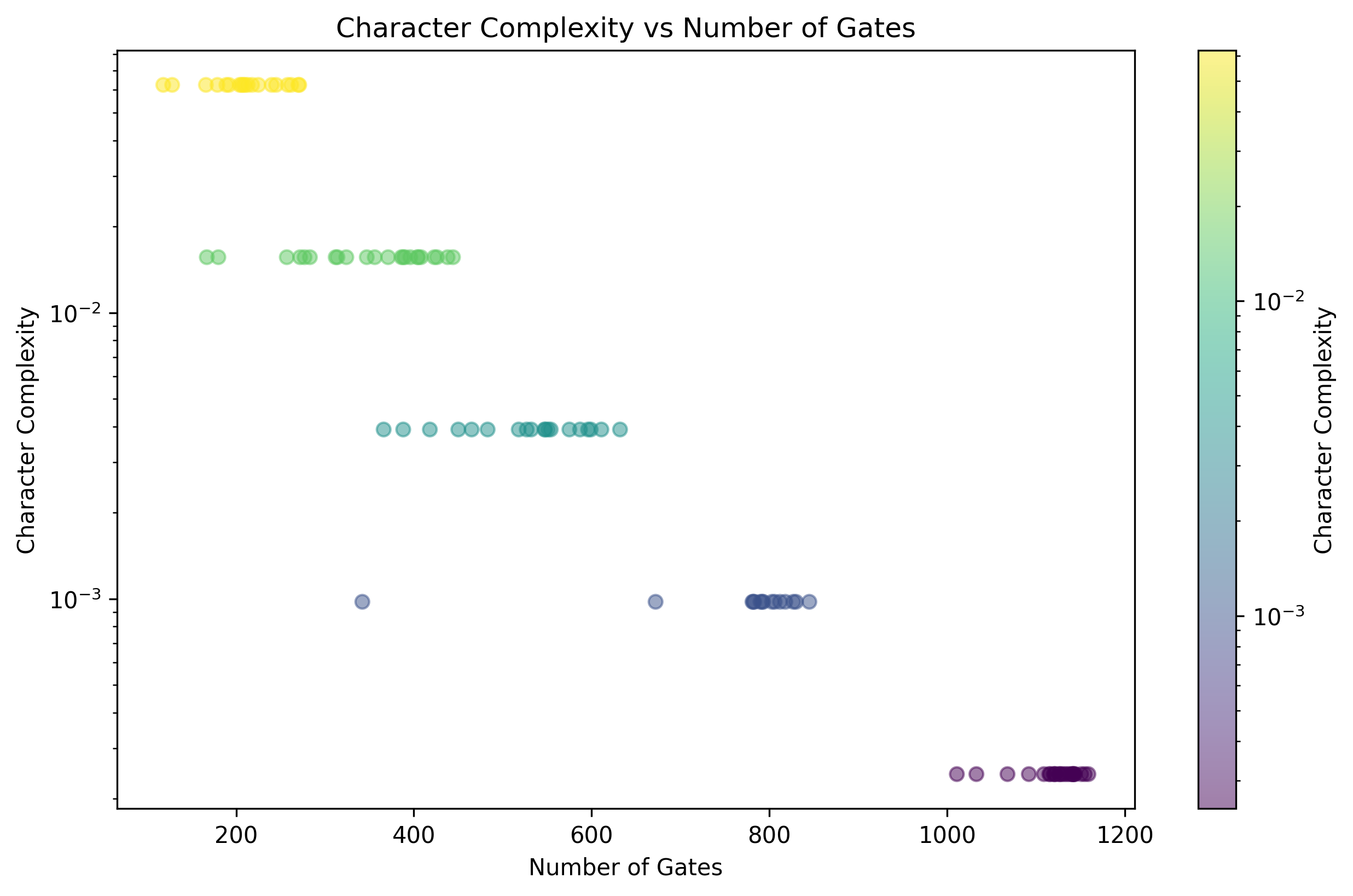}
		\caption{Quantum Approximate Optimization Algorithm}
		\label{fig:qaoa_scaling}
	\end{subfigure}
	\caption{Character Complexity scaling with respect to gate count for different 6-qubit circuit types.}
	\label{fig:complexity_scaling}
\end{figure}

Figure \ref{fig:complexity_scaling} illustrates how character complexity scales with increasing gate count for each circuit type:

\begin{itemize}
	\item \textbf{Random circuits (Figure \ref{fig:random_circuit_scaling})} display a rapid decrease in complexity with increasing gate count. This suggests that adding more gates to a random circuit quickly introduces significant scrambling.
	
	\item \textbf{QFT (Figure \ref{fig:qft_scaling})} shows a more gradual decrease in complexity compared to random circuits. This could reflect the structured nature of the QFT algorithm, where scrambling is introduced more slowly as the circuit grows.
	
	\item \textbf{QAOA (Figure \ref{fig:qaoa_scaling})} exhibits a slower decrease in complexity compared to both random circuits and QFT. This could correspond to QAOA's iterative approach, where each iteration introduces some scrambling while maintaining problem-related structure.
\end{itemize}

These comparisons highlight how character complexity can reveal fundamental differences and similarities in the scrambling behavior of various quantum algorithms. The unique scaling behaviors and complexity distributions observed for different circuit types underscore the diversity of scrambling dynamics in the quantum algorithmic landscape.

Some potential implications and future directions suggested by these results include:

\begin{enumerate}
	\item \textbf{Scrambling and Quantum Advantage}: The relationship between scrambling (as measured by character complexity) and quantum advantage could be investigated. Do algorithms that exhibit more scrambling offer greater potential for speedup over classical methods?
	
	\item \textbf{Optimal Scrambling}: The character complexity profiles could guide the design of algorithms that achieve an optimal balance between scrambling (for computational power) and structure preservation (for specific problem solving).
	
	\item \textbf{Scrambling and Error Correction}: The interplay between scrambling and error correction could be explored. How does the introduction of error correction affect the scrambling behavior of different algorithms?
	
	\item \textbf{Scrambling in Noisy Intermediate-Scale Quantum (NISQ) Algorithms}: Character complexity could be used to study the scrambling behavior of NISQ algorithms, which operate under the constraints of current quantum hardware. This could provide insights into the potential and limitations of these algorithms.
\end{enumerate}

By interpreting character complexity as a measure of scrambling, we gain a new perspective on the behavior and properties of quantum algorithms. This opens up exciting avenues for future research at the intersection of quantum complexity, quantum chaos, and quantum algorithm design.

\section{Discussion} The introduction of character complexity as a novel measure for quantum circuits provides several new insights into the structure and behavior of quantum algorithms. By leveraging the powerful tools of group representation theory, character complexity captures aspects of quantum circuit complexity that are not fully reflected in traditional measures such as gate count or circuit depth.

One of the key insights gained from character complexity analysis is the existence of a tight upper bound on the complexity of a quantum circuit. This bound provides a fundamental limit on the complexity of quantum circuits and could guide the design of optimal quantum algorithms.

Another significant insight is the strong relationship between character complexity and the classical simulability of quantum circuits. The theorem proving that quantum circuits with logarithmically bounded character complexity can be efficiently simulated classically provides a new perspective on the boundary between classical and quantum computation. This could lead to new approaches for identifying quantum algorithms that offer a genuine quantum advantage.

Compared to other quantum circuit complexity measures, character complexity stands out in its ability to capture the structural properties of quantum circuits that arise from their underlying symmetries. While measures like gate count and circuit depth provide valuable information about the size and time complexity of a circuit, they do not fully reflect the circuit's action on different subspaces of the quantum state space. Character complexity, on the other hand, directly quantifies this action, providing a more nuanced view of a circuit's complexity.

Potential applications of character complexity include the design of more efficient quantum algorithms, the identification of classically simulable quantum circuits, and the development of new quantum circuit optimization techniques. By understanding the character complexity of different quantum operations, it may be possible to design quantum circuits that achieve the desired computational results with lower complexity.

However, there are also some limitations to consider. Computing character complexity for large quantum circuits may itself be a computationally intensive task, potentially limiting its practical applicability in some cases. Additionally, while character complexity provides valuable insights, it is not a complete substitute for other complexity measures, and a holistic analysis of quantum circuits should consider multiple complementary measures.

\section{Conclusion and Future Work} In this paper, I have introduced character complexity as a novel measure for analyzing the complexity of quantum circuits. By leveraging the mathematical tools of group representation theory, character complexity provides a new lens through which to understand the structure and behavior of quantum algorithms.

I have proven several key properties of character complexity, including its bounds, its multiplicativity under tensor products, and its submultiplicativity under circuit composition. Notably, I have shown a strong connection between character complexity and the classical simulability of quantum circuits, providing a new perspective on the boundary between classical and quantum computation.

Furthermore, I have introduced new methods for visualizing character complexity, offering intuitive insights into the complexity landscape of quantum circuits. Our empirical results suggest interesting scaling behaviors of character complexity with respect to qubit count and gate count.

Looking forward, there are several exciting avenues for future research. One key direction is to further explore the practical applications of character complexity in quantum algorithm design and optimization. By understanding the character complexity of different quantum operations and circuits, it may be possible to design more efficient quantum algorithms that achieve the desired results with lower complexity.

Another important direction is to further investigate the relationship between character complexity and other aspects of quantum information, such as entanglement, coherence, and quantum error correction. Exploring these connections could lead to new insights into the fundamental properties of quantum systems and their implications for quantum computing.

Additionally, future work could focus on developing efficient methods for computing character complexity for large-scale quantum circuits. This could involve the development of new algorithmic techniques, as well as the exploration of potential approximation methods.

In conclusion, character complexity provides a powerful new tool for analyzing and understanding quantum circuits. By bridging the gap between abstract group-theoretic concepts and practical quantum computing concerns, it opens up new avenues for the design and optimization of quantum algorithms.

\bibliography{manuscript}

\begin{thebibliography}{10}

\bibitem{gottesman1998heisenberg}
Daniel Gottesman.
\newblock The heisenberg representation of quantum computers.
\newblock {\em arXiv preprint quant-ph/9807006}, 1998.

\bibitem{bravyi2016improved}
Sergey Bravyi and David Gosset.
\newblock Improved classical simulation of quantum circuits dominated by
  clifford gates.
\newblock {\em Physical review letters}, 116(25):250501, 2016.

\bibitem{shami2024bridging}
Daksh Shami.
\newblock Bridging classical and quantum: Group-theoretic approach to quantum
  circuit simulation.
\newblock {\em arXiv preprint arXiv:2407.19575}, 2024.

\bibitem{nielsen2010quantum}
Michael~A Nielsen and Isaac~L Chuang.
\newblock {\em Quantum computation and quantum information}.
\newblock Cambridge university press, 2010.

\bibitem{amy2013meet}
Matthew Amy, Dmitri Maslov, Michele Mosca, and Martin Roetteler.
\newblock Meet-in-the-middle algorithm for fast synthesis of depth-optimal
  quantum circuits.
\newblock In {\em Computer-Aided Design (ICCAD), 2013 IEEE/ACM International
  Conference on}, pages 198--204. IEEE, 2013.

\bibitem{cross2019validating}
Andrew~W Cross, Lev~S Bishop, Sarah Sheldon, Paul~D Nation, and Jay~M Gambetta.
\newblock Validating quantum computers using randomized model circuits.
\newblock {\em Physical Review A}, 100(3):032328, 2019.

\bibitem{shende2006synthesis}
Vivek~V Shende, Igor~L Markov, and Stephen~S Bullock.
\newblock Synthesis of quantum-logic circuits.
\newblock {\em IEEE Transactions on Computer-Aided Design of Integrated
  Circuits and Systems}, 25(6):1000--1010, 2006.

\bibitem{kraus2001optimal}
Barbara Kraus and J~Ignacio Cirac.
\newblock Optimal creation of entanglement using a two-qubit gate.
\newblock {\em Physical Review A}, 63(6):062309, 2001.

\bibitem{bravyi2019simulation}
Sergey Bravyi, Dan Browne, Padraic Calpin, Earl Campbell, David Gosset, and
  Mark Howard.
\newblock Simulation of quantum circuits by low-rank stabilizer decompositions.
\newblock {\em Quantum}, 3:181, 2019.

\bibitem{serre1977linear}
Jean-Pierre Serre.
\newblock {\em Linear representations of finite groups}.
\newblock Springer-Verlag, 1977.

\bibitem{fulton2013representation}
William Fulton and Joe Harris.
\newblock {\em Representation theory: a first course}.
\newblock Springer Science \& Business Media, 2013.

\bibitem{isaacs1976character}
I~Martin Isaacs.
\newblock {\em Character theory of finite groups}.
\newblock Academic Press, 1976.

\end{thebibliography}
\bibliographystyle{unsrt}

\end{document}